\documentclass[proof]{pasj00}
\begin{document}
\SetRunningHead{Masada et al. }{Magnetic Reconnection Model for Magnetar Giant Flare}

\title{Solar-type Magnetic Reconnection Model for Magnetar Giant Flare}

\author{Youhei \textsc{Masada}}
\affil{Hinode Science Project, National Astronomical Observatory of Japan,\\ 2-21-1 Osawa, Mitaka, Tokyo 181-8588}
\email{y.masada@nao.ac.jp}
\author{Shigehiro \textsc{Nagataki}}
\affil{Yukawa Institute for Theoretical Physics, Kyoto University,\\ Kyoto 606-8502}
\author{Kazunari \textsc{Shibata}}
\affil{Kwasan and Hida Observatories, Department of Astronomy, Kyoto University,\\ Kyoto 606-8502}
\and
\author{Toshio \sc{Terasawa}}
\affil{Department of Physics, Tokyo Institute of Technology,\\ Tokyo 152-8551}

%


\maketitle

\begin{abstract}
We present a theoretical model describing magnetar giant flares on the basis of solar flare/coronal mass ejection theory. In our model, 
a preflare activity plays a crucial role in driving evaporating flows and supplying baryonic matters into the magnetosphere. The loaded 
baryonic matter, that is  called "prominence", is then gradually uplifted via crustal cracking with maintaining a quasi-force-free equilibrium 
of the magnetosphere. Finally the prominence is erupted by the magnetic pressure force due to the loss of equilibrium triggered by the 
explosive magnetic reconnection. The giant flare should be induced as a final outcome of the prominence eruption accompanied by  
large-scale field reconfigurations. An essential difference between solar and magnetar flares is the control process of their evolutionary dynamics. 
The flaring activity on magnetars is mainly controlled by the radiative process unlike the solar flare governed by the electron conduction. 
It is highly suggestive that our model is accountable for the physical properties of the extraordinary giant flare observed on 2004 December 27 
from SGR1806-20, including the source of baryonic matters loaded in the expanding ejecta observed after the giant burst.
\end{abstract}
\section{Introduction}
There has recently been growing evidence that soft gamma-ray repeaters (SGRs) and anomalous X-ray pulsars (AXPs) are the same 
population of the ultra-strongly magnetized neutron star ($B \gtrsim 10^{14}$ G), so called "Magnetar" (Duncan \& Thompson 1992; 
Harding \& Lai 2006). Activities in these objects are powered by the dissipation of strong magnetic fields unlike rotation-powered 
pulsars and accretion-powered X-ray binaries. Both SGRs and AXPs generally undergo the quiescent phase with persistent X-ray emission 
and the recurrent phase of soft gamma-ray (Mereghetti et al. 2004). Typical luminosities at these two phases are 
$L_{{\rm x}} \simeq 10^{33}$--$10^{35}\ \ {\rm erg\ s^{-1}}$ and $L_{\gamma } \simeq 10^{38}$--$10^{41}\ \ {\rm erg\ s^{-1}}$ 
respectively (Kaspi et al. 2003; Woods \& Thompson 2006). 

Besides the common short burst, the giant flare with enormous energy and long bursting duration is exceptionally observed from SGRs. 
These are the most energetic galactic event currently known ($\simeq 10^{44}$--$10^{47}\ {\rm erg}$). Only three of giant flares have 
been observed: SGR0526-66 on 1979 March 5 (Mazets et al. 1979), SGR1900+14 on 1998 August 27 (Hurley et al. 1999; 
Kouveliotou et al. 1999; Feroci et al. 2001; Tanaka et al. 2007), and SGR1806-20 on 2004 December 27 
(Hurley et al. 2005; Palmer et al. 2005; Terasawa et al. 2005). 

The giant flare from SGR1806-20 is the most recent and energetic one. It is characterized by an ultra-luminous hard spike, with energy 
$\simeq 10^{46}\ {\rm erg}$, lasting $\simeq 0.1\ {\rm s}$, which decays rapidly into a soft pulsating tail lasting hundreds of second. 
The spectrum of the hard spike is fitted by the blackbody radiation with the temperature $\simeq 10^9\ {\rm K} $. In addition, a 
preflare activity with the total energy $\simeq 10^{41}\ {\rm erg}$ and the duration $\simeq 1.0\ {\rm s}$ is detected $142\ {\rm s}$ 
before the main burst. It also shows the single blackbody spectrum with the temperature $\simeq 10^{8}\ {\rm K} $ (Boggs et al. 2007). 

The exceptional event observed in association with the giant flare from SGR1806-20 is an expanding radio emitting ejecta (Taylor et al. 
2005; Cameron et al. 2005; Gaensler et al.2005). The emission properties are well resolved by the synchrotron radiation from the shocked 
baryonic shell with the mass $\gtrsim 10^{24.5}\ {\rm g}$ and the expansion velocity $\simeq 0.4c $ if it is roughly spherical 
(Gelfand et al. 2005; Granot et al. 2006). However, the origin of the baryon-load of the ejecta remains unsettled although it is essential 
for promoting better understanding of flaring activities on magnetars (see Thompson \& Duncan 1995, hereafter TD95; Gelfand et al. 2005; Lyutikov 2006). 

In these situations, the solar flare gives an important prototype context for the astrophysical flaring activity. It is well known that solar flares are also 
accompanied by analogous mass ejection events, that is called "coronal mass ejections (CME)". The physical properties of the solar flare/CME are very 
similar to those of the magnetar flare (Lyutikov 2006). Specifically, the initial spike and subsequent tail emissions associated with magnetar flares are 
naturally reminiscent of impulsive and decay phases of solar flares. In this paper, we construct a theoretical model for the magnetar giant flare on 
the basis of the solar flare/CME model. 

Our paper is organized as follows.  The magnetic reconnection model, which is the underlying theory of the solar flare/CME, is applied to the magnetar 
system for providing a physical basis of our magnetar model in \S~2. In \S~3, we propose a theoretical model describing the magnetar 
giant flare according to a promising solar flare/CME scenario. In \S~4, we discuss the effectiveness of the assumptions 
used in our model and the baryonic evaporation process in the microscopic viewpoint. Finally, we summarize the characteristics of our model in \S~5. 
\section{Magnetic Reconnection Model}
Magnetic reconnection is believed to be a crucial mechanism of the energy release in the solar flare (Parker 1963; Petschek 1964). 
The observational evidence, such as cusp-shaped soft X-ray loops and hard X-ray sources above the loops, supports this model that 
predicts the primary site of the energy release above the soft X-ray loops (Tsuneta et al. 1992; Masuda et al. 1994). The discovery of 
escaping plasmoids from the flare sites is another evidence for this model because it predicts the ejection of plasma from the 
reconnection region (Shibata et al. 1995; Ohyama \& Shibata 1997). 

The released magnetic energy triggered by the magnetic reconnection is converted to the thermal energy in the magnetosphere and then the 
heat conduction drives the evaporation of chromospheric plasma. The discoveries of blue-shifted component of spectral lines and moving 
plasma in X-ray images, confirm the upward motion of the ablated plasma anticipated from the chromospheric evaporation theory 
(Feldman et al. 1980; Culhane et al. 1992; Doschek et al. 1992; Savy 1997). 

Shibata \& Yokoyama (1999; hereafter SY99) propose a magnetic reconnection model of the solar flare taking account of the chromospheric 
evaporation. They point out that the flare temperature is determined by the counterbalance between the reconnection heating and the 
conductive cooling. The coronal density is then controlled by the evaporation cooling of the chromospheric plasma 
which compensates for the conductive heating (see also, Shimojo et al. 2001; Miyagoshi \& Yokoyama 2003). 

The scaling relation obtained from the magnetic reconnection model can explain the observed correlation between the 
emission measure $EM$ and the flare temperature $T$ from solar micro-flares to proto-stellar flares consistently (Feldman et al. 1995; 
Yokoyama \& Shibata 1998, 2001; Aschwanden et al. 2008). This suggests that the underlying physics of the flare would be common to 
various astrophysical systems. We apply the magnetic reconnection model to the mysterious magnetar system for providing the 
physical basis of the flaring activity. 
\subsection{Energetics of Magnetar Flare} 
\subsubsection{Energy Release by the Magnetic Reconnection}
\begin{figure}	
\begin{center}
\FigureFile(120mm, 70mm) {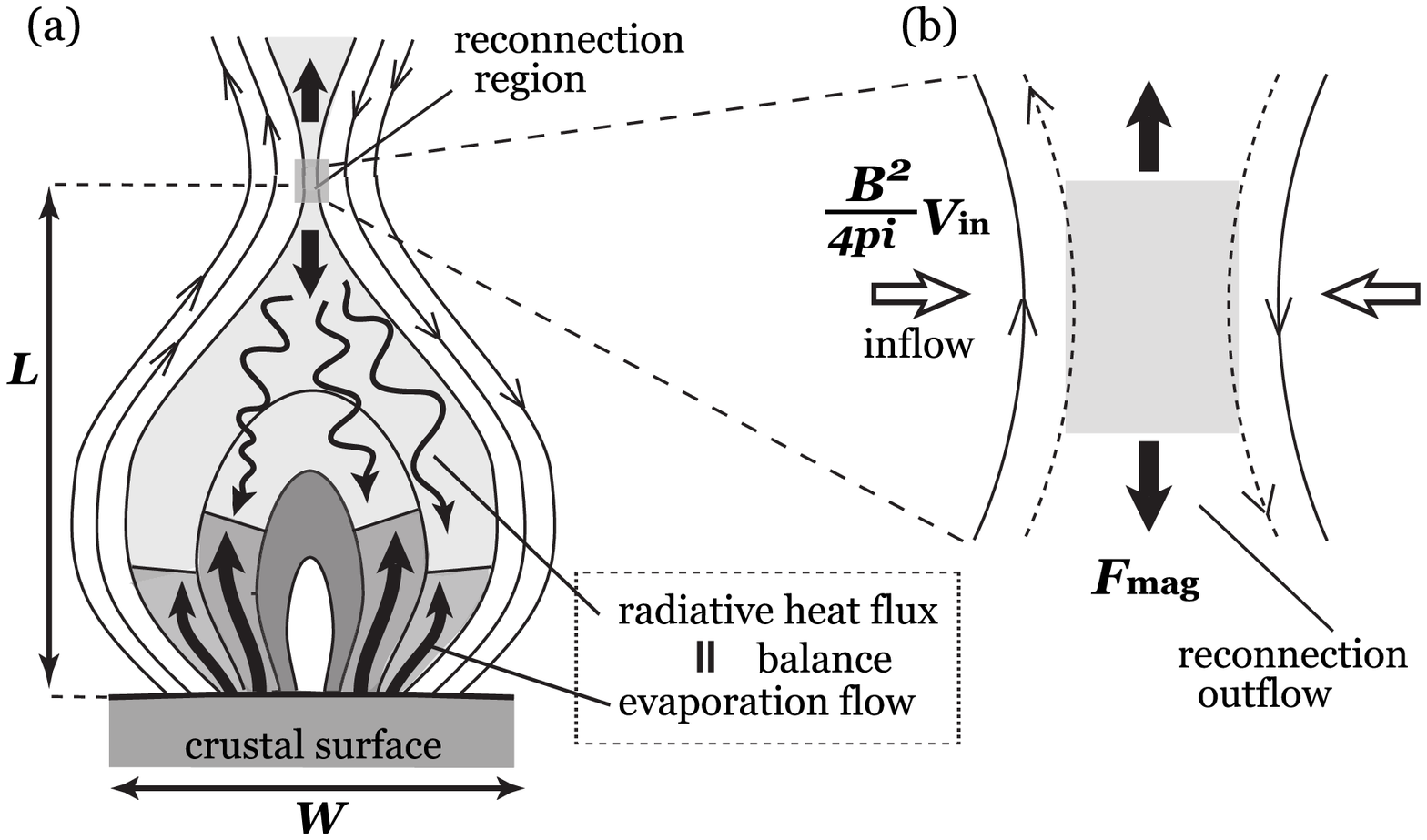}
\end	{center}
\caption{ Panel~(a): A single magnetic arcade loop formed after the magnetic reconnection which triggers the explosive magnetic energy 
release. We suppose now the situation in which a lot of magnetic arcade loops are formed on the magnetar surface. The typical 
reconnection height and width of a single flare loop is represented by $L$ and $W$. The radiative heat flux driven via the magnetic 
reconnection heats the crustal matter and drives the upward evaporation flow into the flare loop. Panel~(b): A schematic view which focuses 
on the reconnection site. The energy release rate by the magnetic reconnection is comparable to the inflow rate of the magnetic energy into 
the reconnection site, that is $E_{\rm flare}/t_{\rm rec} \simeq F_{\rm mag} A = (B^2/4\pi )\ V_{\rm in} A $. } 
\label{fig1}
\end{figure}

We consider a situation in which enormous magnetic energy stored in the magnetosphere is released through a flare induced by the 
magnetic reconnection. Then a lot of magnetic arcade loops are formed on magnetar surface. Here the reconnected single flare loop is schematically 
illustrated in Figure~\ref{fig1}a. Figure~\ref{fig1}b focuses on a reconnection site to clearly specify the energy conversion process. 

At first, we provide a physical process that controls the typical energy liberated by magnetar flares. When we follow a classical 
magnetic reconnection model, the energy release rate of a flare can be described using the released magnetic energy flux $F_{\rm mag}$
\begin{equation} 
\frac{{\rm d} E_{\rm flare}}{{\rm d} t } = F_{\rm mag}\ A  \;, \label{eq1} 
\end{equation}
(Priest \& Forbes 2002) where $E_{\rm flare}$ is the released energy by a flare, and $A$ is the area of the reconnection site. 

As shown in Figure~\ref{fig1}b, the magnetic reconnection can liberate the magnetic energy which is equivalent to that inflows 
into the reconnection site. Given the magnetic energy density $B^2/4\pi $ and the inflow velocity $V_{\rm in}$, 
the magnetic energy flux $F_{\rm mag}$ is 
\begin{equation}
F_{\rm mag} = \frac{B^2}{4\pi} V_{\rm in} \;, \label{eq2}
\end{equation}
where $B$ is the strength of the magnetic field. 

If we consider only a single flare loop, the area of reconnection site would be given by $A = LW$, where $L$ is the the height of the 
reconnection point and $W$ is the width of the single flare loop (see fig.~\ref{fig1}a). However, we now suppose that there are formed 
a lot of magnetic arcade loops on the magnetar surface. Hence for taking account of the contribution from all magnetic loops, we adopt 
$A \simeq LR$ as the total area of all reconnection sites, where $R$ is the size of the active region sustaining flaring activities. We fix the 
typical size $R$ as $10^{6}$ cm in the following. 

Since the reconnection timescale is evaluated as $ t_{\rm rec} \simeq L/V_{\rm in}$, equation~(\ref{eq1}) becomes
\begin{equation}
E_{\rm flare} = \frac{B^2}{4\pi} RL^2\;, \label{eq3}  
\end{equation}
where the approximation ${\rm d}E_{\rm flare}/{\rm d}t \simeq E_{\rm flare}/t_{\rm rec}$ is used for obtaining this equation. 
Based on the classical magnetic reconnection theory, the field strength $B$ and the reconnection height $L$ mainly 
controls the magnetic energy liberated by the flare. 
\subsubsection{Thermal Balance of Magnetar Flare}
In the case of the solar flare, the flare temperature $T$ is determined by the conductive cooling which balances with the reconnection 
heating (SY99). However, the radiative cooling dominates the conductive one in the physical condition realized in the magnetar flare 
(see \S~4.3 in detail). The temperature in the flaring state of the magnetar is thus mainly controlled by the radiative cooling which 
compensates for the reconnection heating. 

Additionally, we should take account of the heat absorption due to the pair creation which would impact on the 
thermal balance of the magnetar system especially in the higher temperature regime $T \lesssim 10^{10}$ K. 
Considering that the blackbody cooling and the heat absorption due to the pair creation become predominate in 
various cooling processes, the thermal balancing equation gives
\begin{eqnarray}
E_{\rm flare} & = & 4\pi R^2 c \Delta t \ (U_{\rm rad} + U_{e^\pm}) \nonumber \\
& = & 4\pi R^2 c \Delta t \left[ \sigma_B T^4/c  + m_e n_{\pm } c^2\right] \;. \label{eq4}
\end{eqnarray}
where $\Delta t$ is the flare duration, $\sigma_B$ is the Stefan-Boltzmann constant, $c$ is the speed 
of light, $m_e$ is the electron mass, and $n_{\pm}$ is the density of the electron positron pair. Note that $U_{\rm rad}$ represents the 
radiative energy due to the blackbody cooling and $U_{e^{\pm}}$ shows the endothermic energy due to the pair creation. The magnetic 
effects on the thermal equilibrium are expected to be vanishingly small and ignored here because the configuration of post-flare loops 
would not be changed by post-reconnection processes. 

We focus on the system with $B = 10^{15}\; {\rm G}$ and $T \ll 10^{10}\; {\rm K}$ which are suitable conditions for describing 
the flaring activity of the magnetar (Boggs et al. 2007). Assuming the local thermal equilibrium (LTE), in the range $B \gg B_{\rm QED}$ and 
$T \ll m_ec^2/k_B \simeq 10^{10}\; {\rm K}$, the density of the electron positron pair supplied by pair creations is
\begin{equation}
n_{\pm} \equiv \frac{(m_e c)^3}{\hbar^3(2\pi^3)^{1/2}} \left( \frac{B}{B_{\rm QED}}\right) 
\left( \frac{ k_B T}{m_e c^2} \right)^{1/2} \exp \left( - \frac{m_e c^2}{k_B T} \right) \;, \label{eq5}
\end{equation}
[see eq.~(75) in Thompson \& Duncan 2001], $h = 2\pi\hbar $ is the Planck constant, $k_B$ is the Boltzmann constant and 
$B_{\rm QED} = m_e^2 c^3/e\hbar = 4.4 \times 10^{13} \ {\rm G}$ is the magnetic flux density at which the energy of the first electron 
Landau level becomes comparable to the electron rest mass. By solving the nonlinear equation~(\ref{eq4}) coupled with the 
equation~(\ref{eq3}) and (\ref{eq5}), we can derive the flare temperature as a function of four physical parameters, 
$T = T (L,B,R,\Delta t)$ or $T (E_{\rm flare},B,R,\Delta t )$. 

\subsection{Baryonic Evaporation during Magnetar Flare}
Once the solar flare begins, the released energy is rapidly transported to the top of the chromosphere by the electron heat conduction 
and heats the chromospheric plasma suddenly. Then the pressure of the heated plasma increases drastically and drives the upward flow 
into the magnetic loop. A hot post-flare loop, which is filled by the evaporated dense plasma, should be formed finally 
(Hirayama 1974; Sylwester 1996; SY99). It can be anticipated to operate a similar baryonic evaporation process in the magnetar flare 
(see Liu et al. 2002 for an application of chromospheric evaporation process to accretion disks). 

In the magnetar flare, the photon flux plays a crucial role in the heat transport unlike the solar flare dominated by the electron heat 
conduction. This is because the mean free path of the electron is very short and it thermalizes instantaneously in the magnetar's magnetosphere. 
Even if the energy is transported by the created electron-positron pair, the evaporation eventually occurs since the 
thermal equilibrium should be established by the photon flux after the pair beams are thermalized. 

The incident energy flux, which inflows into the crustal surface, should thus counterbalance with the outgoing enthalpy flux of the 
evaporation flow (see fig.~\ref{fig1}a). When we consider the contribution from the created electron positron pair during the flare, 
the number density of the baryon in the evaporation flow $n_{\rm ev}$ is provided by the balancing equation; 
\begin{equation}
F_{\rm heat}  =  ( h + h_{\pm} ) v_{\rm ev} \;,  \label{eq6} 
\end{equation}
(c.f., SY99) where $F_{\rm heat } = E_{\rm flare}/(4\pi R^2\Delta t) $ is the inflowing downward energy flux, $v_{\rm ev}$ is the upward 
velocity of the evaporation flow, $h$ is the specific enthalpy of composite gas of the baryon and the equilibrium radiation field, and 
$h_{\pm}$ is that of the created electron positron pair. Notice that this equation is equivalent to the energy conservation equation 
for a mass conserving steady system with the composite gas and the created electron positron pair (c.f., Mihalas \& Mihalas 1984). 

Assuming the thermal equilibrium state, the specific enthalpy of the composite gas $h$ is 
\begin{eqnarray} 
h & = & n_{\rm ev} k_{B} T \left( \frac{5}{2} + 4\alpha \right)  \;, \label{eq7} \\
\alpha & \equiv & \frac{P_{\rm rad}}{P_{\rm gas}} = \frac{4\sigma_{B}T^3}{3 c n_{\rm ev}  k_{B}} \;, \label{eq8} 
\end{eqnarray} 
(Mihalas \& Mihalas 1984) where $n_{\rm ev}$ is the number density of the baryon loaded in the evaporation flow, 
and $\alpha $ is the ratio of radiation and gas pressures. The specific enthalpy of the electron positron pair $h_{\pm}$ 
is additionally given by 
\begin{equation}
h_{\pm } = n_{\pm } m_e c^2 + \frac{5}{2} n_{\pm} k_B T  \;, \label{eq9} 
\end{equation} 
where $n_{\pm}$ is the pair density given by equation~(\ref{eq5}).  

The upward velocity of the evaporation flow $v_{\rm ev}$ can be replaced by the sound speed of the composite gas $C_s$, 
according to the chromospheric evaporation theory of the sun (SY99; Shimojo et al. 2001),
\begin{eqnarray}
v_{\rm ev} & \simeq & C_{s} = c\left[ \frac{\Gamma n_{\rm ev}  k_{B} T_{\rm flare} (1 + \alpha )}{ n_{\rm ev} m_{p} c^2 + h }\right]^{1/2}  
\;, \label{eq10} \\
\Gamma &\equiv & \frac{ 5/2 + 20\alpha + 16\alpha^2}{ ( 3/2 + 12\alpha )( 1+\alpha ) } \;, \label{eq11}
\end{eqnarray}
(Mihalas \& Mihalas 1984) where $m_p $ is the proton mass. We thus find from the equation~(\ref{eq6}) that the baryon density of the
evaporation flow is the function of $E_{\rm flare}$, $B$, $R$, $\Delta t$ since the flare temperature is given 
by $T = T\ (E_{\rm flare},B,R,\Delta t )$.  

Note that it is not settled whether the significant downward energy flux can be maintained after the radiation and pair energy density 
sufficiently grows at the surface in the realistic magnetar situation. We naively assume here that  it is sustained as long as the flaring 
activity lasts. The duration sustaining the strong downward photon flux will be investigated in our future numerical work. 
\subsection{Numerical Solutions Describing the Evaporation}
For given parameters $E_{\rm flare}$, $B$, $R$ and $\Delta t$, we can obtain $T$, $v_{\rm ev}$, $\alpha $, $n_{\pm}$ and $n_{\rm ev}$ by 
iteratively solving the coupled equations~(\ref{eq4})--(\ref{eq11}). The evaporated baryonic mass $M_{\rm ev}$ is derived from a 
relation $M_{\rm ev} = 4\pi R^2 m_p n_{\rm ev} v_{\rm ev} \Delta t $ (Shimojo et al. 2001). Note that the size of the active region 
and the field strength are fixed as $R = 10^6\ {\rm cm}$ and $B = 10^{15}\ {\rm G}$ respectively. 

\begin{figure}[t]
\begin{center}
\begin{tabular}{cc}
\FigureFile(80mm, 50mm){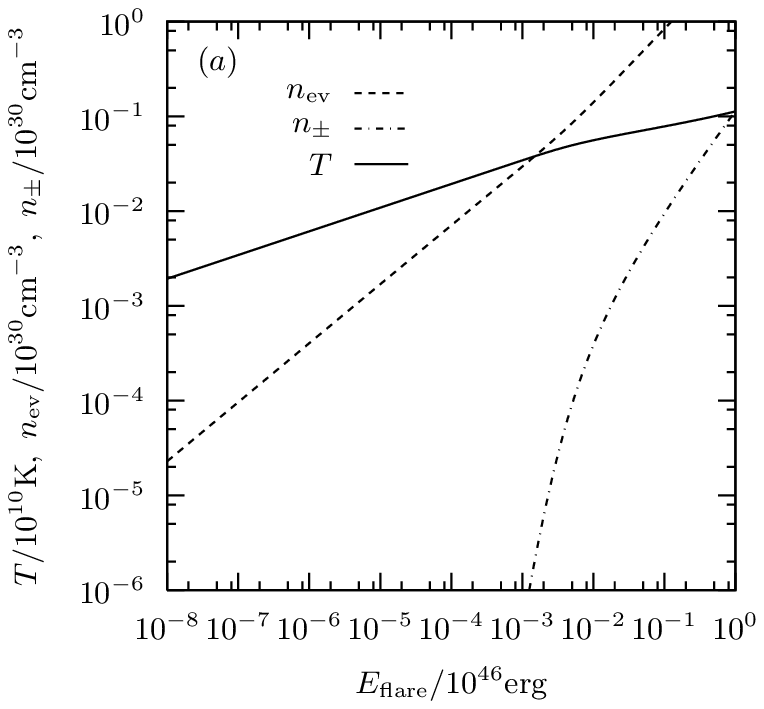} &
\FigureFile(80mm, 50mm){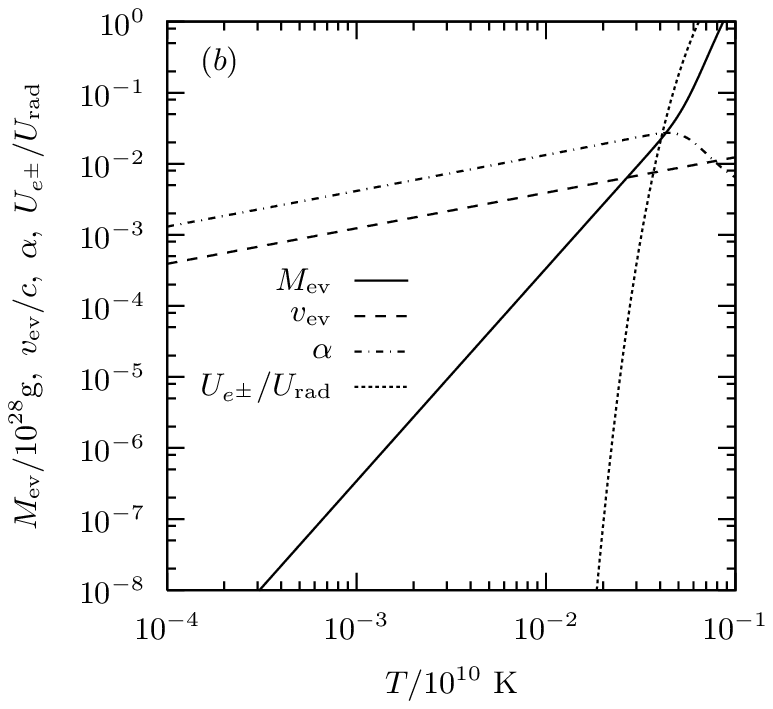} 
\end{tabular}
\end{center}
\caption{(a): The flare temperature $T$, the evaporated baryon density $n_{\rm ev}$ and the created electron positron pair density 
$n_{\pm }$ as a function of an arbitrary parameter $E_{\rm flare}$ in the case with the fixed flare duration $\Delta t = 1.0\ {\rm s}$. 
(b): The mass of the evaporated baryon $M_{\rm ev}$, the evaporation velocity $v_{\rm ev} $, the ratio of the radiation and gas pressures 
$\alpha$ and the ratio of energy densities for the electron positron pair and the radiation field $U_{e^{\pm}}/U_{\rm rad}$ as a function 
of the flare temperature $T\: [\: = T(E_{\rm flare},B,R,\Delta t)]$ in the case with $\Delta t = 1.0\ {\rm s}$. The size of the active 
region and the field strength are fixed as $R= 10^6\; {\rm cm}$ and $B = 10^{15}\; {\rm G}$ in these two charts. Note that all 
physical variables are normalized by their typical values. } 
\label{fig2}
\end{figure}

Figure~\ref{fig2}a shows the flare temperature $T$, number densities of the evaporated baryon $n_{\rm ev}$ and the created electron 
positron pair $n_\pm $ as a function of an arbitrary parameter $E_{\rm flare}$ in the case with the fixed flare duration 
$\Delta t = 1.0\ {\rm sec}$. The each variables are normalized by their typical values. 

As is expected, all physical variables increase with increasing the flare energy. Notice that the growth rate of the flare temperature is 
slightly reduced in the high energy range $E_{\rm flare} \gtrsim 10^{44}\ {\rm erg}$ by the cooling effect due to the pair creation. 
However, the baryon density steadily increases because the enthalpy of the created pair is negligible compared with that of the baryon 
in the range of interest. 

The mass of the evaporated baryon $M_{\rm ev}$, the evaporation velocity $v_{\rm ev}$, the ratio of radiation and gas pressures $\alpha$, 
and the ratio of energy densities for the electron positron pair and the radiation field $U_{e^\pm}/U_{\rm rad}$ are demonstrated as a 
function of the flare temperature $T$ $[\:=\: T\ (E_{\rm flare}, B, R, \Delta t)]$ respectively in Figure~\ref{fig2}b. The flare 
duration is fixed as $\Delta t = 1.0\ {\rm sec}$ again. 

It is found from this figure that the ratio of energy densities $U_{e^\pm}/U_{\rm rad}$ increases with the increase of the flare 
temperature, and reaches to unity around $T \simeq 10^{9}$ K. This indicates that the pair production cooling partially contributes to 
the thermal balance of the system in the range $T \ll 10^{10} $ K. 

Furthermore, the ratio of the radiation and gas pressures $\alpha $ slightly decreases in the high temperature range. This is because 
the increasing rate of the radiation pressure is suppressed due to the pair creation although the gas pressure increases steadily when 
$T \gtrsim 10^{8}$ K. The apparent enhancement of the growth rate for the baryonic mass $M_{\rm ev}$ reflects the reduction of the 
temperature increasing rate due to the pair production cooling.

\begin{figure}[htbp]
\begin{center}
\FigureFile(80mm, 50mm){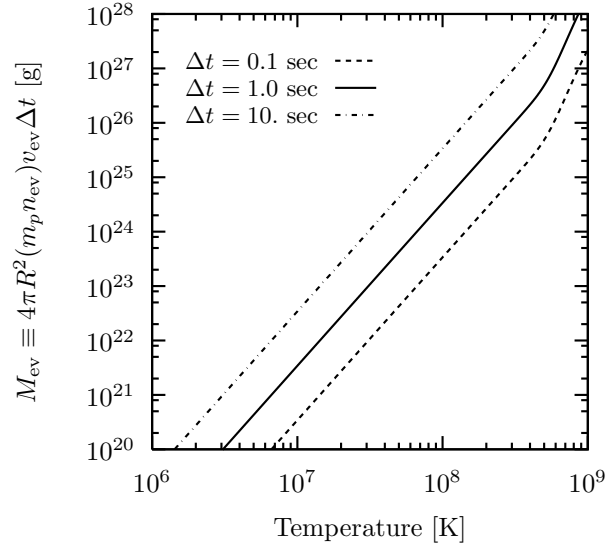}
\end{center}
\caption{ The mass of the evaporated baryon $M_{\rm ev}$ as a function of the flare temperature $T$ in the cases with different flare 
durations $\Delta t = 0.1$, $1.0$ and $10.0\ {\rm sec}$. The size of the active region and the field strength are fixed as 
$R= 10^6\; {\rm cm}$ and $B = 10^{15}\; {\rm G}$ here. The mass of the evaporated baryon becomes larger when the longer flare duration. }
\label{fig3}
\end{figure}

We finally depict the evaporated baryonic mass $M_{\rm ev}$ in Figure~\ref{fig3} as a function of the flare temperature $T$ in the cases 
with different flare durations $\Delta t = 0.1$, $1.0$, and $10.0\; {\rm sec}$. We find that the evaporated baryonic mass is definitely larger 
in the case with longer flare duration. The characteristics of curves are almost same in three models. Numerical results indicate that the 
pair production effect does not dramatically change the qualitative features of our model in the temperature range $T \ll 10^{10} $K. 

We can thus neglect the contributions from the radiation field and the created pair plasma in the evaporation process 
described by equation~(\ref{eq6}). Since the thermal balance of the system can be also retained only by the blackbody 
cooling in equation~(\ref{eq4}), we can analytically derive the flare temperature and the evaporated baryonic mass using scaling relations,
\begin{equation}
T \simeq (16\pi^2 R \sigma_B \Delta t)^{-1/4} B^{1/2} L^{1/2} \;, \label{eq12}
\end{equation}
and 
\begin{eqnarray}
M_{\rm ev}   & \simeq &   \frac{(64\pi )^{1/4}}{5} \sigma_B^{1/4}k_B^{-1}m_p R^{1/2} E_{\rm flare}^{3/4}\Delta t^{1/4} \;,  \nonumber \\
            & \simeq &   \frac{8\pi}{5} \sigma_B k_B^{-1} m_p R^2 T^3\Delta t\;. \label{eq13} 
\end{eqnarray}
These can be used only when the flare energy $E_{\rm flare} \ll 10^{48}\ {\rm erg}$ and the flare temperature $T \ll 10^{10}$K. 
\section{Solar-type Magnetic Reconnection Model for Magnetar Giant Flare}
A promising solar flare/CME scenario, which is strongly supported by observational and theoretical studies  (e.g., Priest \& Forbes 2002; 
Shibata 2005), indicates that the baryonic material loaded in the CME is evaporated  from the sub-coronal chromospheric region before its 
erupting stage. The evaporated matter is then trapped in the coronal region and held in the mechanical equilibrium retained by the balance 
between the magnetic tension and magnetic pressure forces. The CME event is finally driven by the magnetic pressure force after the loss of equilibrium 
which is caused by the dissipation of the magnetic tension via the magnetic reconnection. The CME induces large-scale field reconfigurations 
and triggers the main bursting activity as the final outcome. 

\begin{figure}[htbp]
\begin{center}
\FigureFile(160mm, 100mm){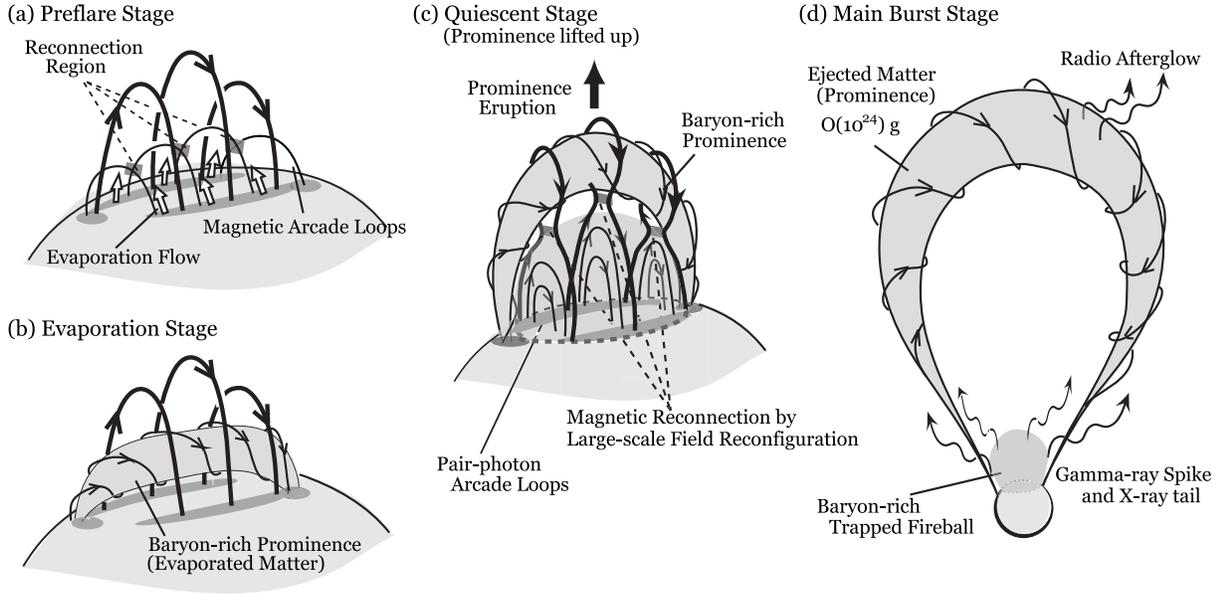}
\end{center}
\caption{A theoretical model for magnetar giant flares on the basis of the solar-type magnetic reconnection theory: (a). At preflare stage, 
enormous magnetic energies stored in the magnetosphere is partially liberated by the magnetic reconnection. The radiative 
heat flux transports the liberated energy and heats the magnetar's crust. The pressure of crustal matters then increases suddenly and 
drives evaporating flows. (b). A baryon-rich prominence with a mass $M_{\rm pro} \sim 10^{24}\ {\rm g} $ is builded up 
due to the mass evaporation in the lower part of the magnetosphere [$\sim O(10^2)$ cm].  (c). The prominence is gradually uplifted 
by the magnetic energy supplied from beneath the magnetar surface via crustal cracking during quiescent stage. Finally, the prominence eruption, 
which is initiated by the loss of equilibrium triggered by the magnetic reconnection, induces large-scale field reconfigurations and triggers 
a giant burst. (d). The liberated energy at the main burst stage is converted to the kinetic energy of the erupted prominence and the radiative energy 
of trapped fireball. The ejected prominence would be observed as radio emitting ejecta and the trapped fireball is expected to 
be the origin of $\gamma$-ray spike followed by pulsating hard X-ray tails. }
\label{fig4}
\end{figure}

According to the solar flare/CME scenario, we propose a model describing magnetar giant flares on the basis of the underlying magnetic 
reconnection theory constructed in \S~2. Our model consists of the following four stages which are illustrated in Figure~\ref{fig4} schematically. 

(a). A flaring activity begins from preflare stage. In this stage, the magnetic energy stored in the magnetar's magnetosphere is partially 
liberated by the magnetic reconnection (Fig~\ref{fig4}a). Supposing that the typical height of the reconnection point is relatively 
low and is given by an order of $L = 10^3\ {\rm cm}$, the released energy and the temperature characterizing the preflare, 
$E_{\rm pre}$ and $T_{\rm pre}$, are evaluated from equations~(\ref{eq3}) and~(\ref{eq12}), 
\begin{eqnarray}
E_{\rm pre} & = & 8.0\times 10^{40} B_{15}^2 R_6L_{3}^2 \ \ {\rm erg}\;, \label{eq14} \\
T_{\rm pre} & = & 1.0\times 10^8 B_{15}^{1/2} R_6^{-1/4} L_3^{1/2}\Delta t_{0}^{-1/4} \ \ {\rm K} \;, \label{eq15}
\end{eqnarray}
where $B_{15}$, $\Delta t_{0}$, $R_6$ and $L_3 $ are the field strength, the flare duration, the size of the active region, and the 
reconnection height in units of $10^{15}\ {\rm G} $, $1.0\ {\rm sec}$, $10^6\ {\rm cm} $, and $10^3\ {\rm cm} $ respectively 
(c.f., Boggs et al. 2007). 

(b). During the preflare activity, the radiative heat flux transports the released energy and heats the crustal sub-surface matter. The pressure of 
the crustal matter then increases drastically and drives the upward evaporation flow. As the result, a hot and dense prominence, which is trapped by the 
post-flare loops, is builded up (fig.~\ref{fig4}b). The mass of the prominence $M_{\rm pro}$ is comparable to that of baryonic matters evaporated and is 
given by equation~(\ref{eq13}) 
\begin{equation}
M_{\rm pro} = 3.4 \times 10^{24} R_6^2 T_8^3 \Delta t_0 \ \ {\rm g}\;, \label{eq16}
\end{equation} 
where $T_8 $ is the normalized preflare temperature $ T/10^8\ {\rm K}$ [see the typical preflare temperature in 
eq.~(\ref{eq15})]. 

Immediately after the preflare stage, the formed prominence is bound to the lower magnetosphere 
of the height $ \sim O(100)$ cm gravitationally, while the reconnection point where the preflare is triggered is an 
order of $10^3$ cm [see eq.(14)]. This is because the potential energy of the evaporated matter becomes restricted by the 
input radiative energy (liberated magnetic energy). 

(c). After the preflare stage, the prominence is gradually lifted up by the magnetic energy injected from the magnetar's interior during quiescent stage 
(see \S~4.1 for details of the prominence uplifting process). The system evolves with retaining a quasi-force-free equilibrium by counterbalancing the 
magnetic tension with the magnetic pressure. The prominence is finally erupted by the magnetic pressure force after the loss of equilibrium which is 
caused by the dissipation of the magnetic tension via the magnetic reconnection. The prominence eruption induces large-scale field 
reconfigurations and triggers a giant burst as the final outcome (Fig~\ref{fig4}c). Supposing the reconnection height for liberating numerous magnetic 
energy at the main burst stage to be $L \simeq 4\times 10^{5}\ {\rm cm}$, the energy and temperature of the giant burst $E_{\rm main}$ and $T_{\rm main}$ 
are, from equations~(\ref{eq3}) and (\ref{eq12})
\begin{eqnarray}
E_{\rm main} & = & 1.3\times 10^{46} B_{15}^2 R_6L_5^2 \ \ {\rm erg}\;, \label{eq17} \\ 
T_{\rm main} & = & 3.7\times 10^9 B_{15}^{1/2} R_6^{-1/4} L_5^{1/2}\Delta t_{-1}^{-1/4} \ \ {\rm K}\;, \label{eq18} 
\end{eqnarray}
where $L_5 = L/(4\times 10^5\ {\rm cm})$ and $\Delta t_{-1} = \Delta t /10^{-1}\ {\rm s}$ (c.f., Hurley et al. 2005; Terasawa et al. 2005). 

(d). The released energy at the main burst stage should be converted into the kinetic energy of the erupted 
prominence and the radiative energy of the remained flare loops. The ejected baryon-rich prominence, accelerated by the main burst, 
would be observed as an expanding ejecta (Fig~\ref{fig4}d). We would like to emphasize again that the baryonic matter loaded in the ejecta 
is supplied by the evaporation at the preflare stage, that is $M_{\rm ej} = M_{\rm pro} \simeq 10^{24}\ {\rm g}$. This is consistent 
with the observed value of the baryon load in the radio emitting ejecta in association with the giant flare from SGR1806-20 (c.f., 
Gelfand et al. 2005; Granot et al. 2005). 

The remained flare loops are, in contrast, polluted again by the baryonic matter evaporated by the main burst. The mass of the evaporated 
baryon in this stage $M_{\rm main}$ is also given by equation~(\ref{eq13})
\begin{equation}
M_{\rm main} = 1.7 \times 10^{28} R_{6} T^3_9 \Delta t_{-1} \ {\rm g} \;. \label{eq19}
\end{equation}
where $T_9$ is the temperature at the main burst stage normalized by $3.7\times 10^9 \; {\rm K}$ [see the reference value of 
eq.~(\ref{eq18})]. 

As will be discussed in \S 4.2, the evaporated baryonic matter at the main bust stage would be gravitationally trapped to the magnetar 
surface. The baryon-rich dense flare loops would be the origin of a trapped fireball, and which eventually produce a 
luminous $\gamma$-ray spike and subsequent hard X-ray pulsating tails as is observed in the giant flare from SGR 1806-20. 

Our solar-type magnetic reconnection model is accountable for the flaring activity associated with the giant flare on 2004 December 27 
from SGR 1806-20 consistently. An important prediction from our model is that the preflare activity plays a crucial role in supplying the 
baryonic matter into the potential ejecta "prominence". This suggests that the radio afterglow is expected to be observed only after the giant flare 
with preflare activities, such as the giant burst of SGR 1806-20. 

We would like to stress that the mechanism for baryonic eruptions proposed in our model is magnetic pressure-driven one which is caused by 
the loss of equilibrium triggered by the magnetic reconnection. The preflare-induced mass evaporation plays a role in supplying the baryonic matter 
into the magnetospheric region. The mass ejecting mechanism in our model is thus essentially different from the magnetic tension-driven 
model via slingshot like process proposed by Gelfand et al. (2005). 
\section{Discussion}
\subsection{A Possible Process for Uplifting the Prominence}
We discuss a possible physical mechanism for uplifting the prominence during the quiescent stage in the magnetar system.  
In our model,  we suppose that the magnetic energy required for uplifting the prominence is supplied 
from the magnetar's interior via the crustal cracking by the Lorentz force like as the model proposed by Lyutikov (2006). 
This is because the dipole field of the magnetar is strong enough to deform the neutron star crust. 

During the quiescent stage, the crustal surface would be deformed by the Lorentz force and the magnetic energy 
stored in the magnetar's interior is converted to the motional energy of the crust. The crustal deformation induces the 
twisting of magnetic fields attached to the magnetar surface and generates helical field components. 
The magnetic energy and flux injected into the magnetosphere then re-configure the magnetospheric field. The 
evaporated baryonic matter ($=$ prominence), which is gravitationally bound to the magnetar surface just after the 
preflare stage, would be lifted upwardly in association with the field reconfiguration in the magnetosphere. 

According to Lyutikov (2006), when we consider a crustal plate of size $R$ rotating under the influence of the Lorentz force, 
balanced by viscous stress at the base of the curst, the dissipated energy by the crustal cracking can be evaluated as 
$\sim 10^{44}R_6^4\ \rho_{14} (T_{\rm rot}/0.1\ {\rm sec})^{-1.5}\ {\rm [erg]}$, where $\rho_{14} $ is the density 
normalized by $10^{14}\ {\rm g\ cm^{-3}}$ and $T_{\rm rot} $ is the rotation period of the deep crustal plate (Landau \& Lifshitz 1975). 
Note that the dissipation energy depends on the rotation period of the crustal plate $T_{\rm rot}$. 

Although the relation between the typical deformation time $T_{\rm rot}$ of the deep crust and the duration of the quiescent stage 
is not settled clearly yet, the crustal cracking by the Lorentz force should be a promising mechanism, which is alternative to the buoyant 
flux emergence in the case of the sun, for supplying the magnetic energy into the magnetosphere.  

The process for lifting the prominence remains largely speculative. It is our future work to clarify the physical 
process for triggering the prominence eruption in the magnetar system. For verifying the validity of our model, we are now working on 
the systematic study of magnetar's flaring activities using relativistic MHD simulations (Matsumoto et al. 2010 submitted). 
\subsection{Suitable Stage for the Baryon Loading}
We discuss the suitable stage for supplying the baryonic matter into the potential ejecta. There are two candidate stages, one is 
the preflare stage and the other is the main burst stage. From equations~(\ref{eq16}) and (\ref{eq19}), the baryonic masses evaporated
during each stages are 
\begin{eqnarray}
&&M_{\rm pre}    \simeq  3.4 \times 10^{24} T_{8}^{3} R_{6}^{2} \Delta t_{0}  \; {\rm g} \;,  \label{eq20} \\
&&M_{\rm main}   \simeq  1.7 \times 10^{28} T_{9}^{3} R_{6}^{2} \Delta t_{-1} \; {\rm g} \;,  \label{eq21} 
\end{eqnarray}
where $M_{\rm pre} $ is the evaporated mass during the preflare stage which is comparable to $M_{\rm pro}$. 
The observational constraint on baryonic mass loaded in the ejecta from SGR 1806-20 is $M_{\rm ej} \gtrsim 10^{24.5}$ g 
(Gelfand et al. 2005; Granot et al. 2006). Both stages can supply the sufficient baryonic matter satisfying the constraint.

On the other hand, the gravitational binding energies of the evaporated matters are, at each stages, 
\begin{eqnarray}
E_{g,\rm pre} & = & GM_{\rm NS}M_{\rm pre}/R_{\rm NS}  \nonumber \\
&\simeq&  6.8 \times 10^{44} T_{8}^{3} R_{6}^2 \Delta t_{0} 
\; {\rm erg} \;,  \label{eq22} \\
E_{g,\rm main} & =  & GM_{\rm NS}M_{\rm main}/R_{\rm NS} \nonumber \\
&\simeq& 3.4 \times 10^{48} T_{9}^{3} R_{6}^2 \Delta t_{-1} 
\; {\rm erg} \;,  \label{eq23} 
\end{eqnarray}
where $E_{g,{\rm pre}}$ is the binding energy of the baryonic matter supplied during the preflare stage, $E_{g,{\rm main}}$ is that during 
the main burst stage, $G $ is the gravity constant, $M_{\rm NS} = 1.5 M_\odot$ is the mass of the neutron star, and $R_{\rm NS}$ is the radius 
of the neutron star given by $10^{6}\; {\rm cm}$. 

It is found that the binding energy of the evaporated matter supplied during the main burst stage is much larger than 
the bursting energy of the giant flare given by equation~(\ref{eq17}), that is $E_{\rm main} \simeq 10^{46}\; {\rm erg} \ll E_{g, \rm main}$. 
The baryon supplied by the main burst should be trapped on the magnetar surface without escaping. 
On the other hand, the preflare-supplied baryonic matter can escape, by accelerating the giant burst, from the gravitational field of 
the magnetar. It should be thus the preflare activity that supplies the baryonic matter into the potential ejecta. 
\subsection{The Process Sustaining the Thermal Equilibrium }
\subsubsection{Optically Thick Flare Loop}
In our model, we naively assume that the blackbody cooling mainly retains the thermal balance 
of the system. Here we validate the effectiveness of this assumption. 
At first, the optical thickness of post-flare loops should be examined to check the availability of the blackbody.

We can obtain the number densities of the evaporated baryon during the preflare and main burst stages $n_{\rm pre}$ and $n_{\rm main}$ 
from equations~(\ref{eq20}) and (\ref{eq21})
\begin{eqnarray}
n_{\rm pre}  & \simeq & M_{\rm pre}/(m_p LR^2)  \nonumber \\
& = & 2.1\times 10^{33} L_{3}^{-1} T_{8}^3 \Delta t_{0}  \ {\rm cm^{-3}} \;, \label{eq24} \\
n_{\rm main} & \simeq & M_{\rm main}/(m_p LR^2) \nonumber \\
& = & 1.1\times 10^{34} L_{6}^{-1} T_{9}^3 \Delta t_{-1} \ {\rm cm^{-3}} \;. \label{eq25} 
\end{eqnarray}
The Rosseland mean scattering cross-section in the direction parallel to the magnetic field is 
\begin{equation} 
\sigma_{\rm es}  = 2.2\times 10^9 T^2 B^{-2} \sigma_{T}  \;, \label{eq26}
\end{equation} 
(Silant\'ev \& Yakovlev 1980) where $\sigma_T$ is the Thomson scattering cross-section defined by $(8\pi /3)(e^2/m_e c^2)^2$. 
Using equations~(\ref{eq24})--(\ref{eq26}), the optical depths of post-flare loops are given, at each stages, 
\begin{eqnarray}
\tau_{\rm pre}  & = & n_{\rm pre } \sigma_{\rm es} L \nonumber \\
&\simeq& 3.1 \times 10^{7 } T_{8}^5 B_{15}^{-2} \Delta t_{0 }  \;, \label{eq27}  \\
\tau_{\rm main} & = & n_{\rm main} \sigma_{\rm es} L \nonumber \\
& \simeq & 2.1 \times 10^{14} T_{9}^5 B_{15}^{-2} \Delta t_{-1}  \;. \label{eq28}  
\end{eqnarray}
These indicate that the post-flare loops are optically dense and can be treated as the blackbody sources. 
\subsubsection{Dominant Cooling Process}
Using physical parameters describing the preflare stage, we compare energy evacuation rates in various cooling processes and confirm that the blackbody 
cooling plays a main role in retaining the thermal balance of the system. The cooling rate sustained by the blackbody radiation from optically thick post-flare loops is 
\begin{eqnarray}
\Lambda_{\rm bb} & = & \sigma_B T^4 \nonumber \\
&\simeq& 5.7  \times 10^{27} T_{8}^4 \ \ {\rm erg\ cm^{-2}sec^{-1}} \;. \label{eq29} 
\end{eqnarray}

There are two other cooling processes expected in the magnetar's magnetosphere. One is the radiative heat conduction $\Lambda_{r}$ and 
the other is the electron heat conduction $\Lambda_{e}$ (TD95). The cooling rates by each conductive processes are
\begin{eqnarray}
\Lambda_{r} & = & \kappa_r {\rm d} T/{\rm d}z  \nonumber \\
& \simeq &1.4 \times 10^{21} T_{8}^{-1} B_{15}^2 L_3^{-1}\Delta t_0 \ {\rm erg\: cm^{-2}\: sec^{-1}} \;, \label{eq30} \\  
\Lambda_{e}  & = & \kappa_e {\rm d} T/{\rm d}z  \nonumber \\
&\simeq& 1.0\times 10^{19} T_{8}^{7/2} L_3^{-1}\ {\rm erg\: cm^{-2}\: sec^{-1}} \;, \label{eq31} 
\end{eqnarray}
where 
\begin{equation}
\kappa_{r} \equiv  16\sigma_B T^3/(3Y_e n \sigma_{\rm es}) \;,\ \ \ \  \kappa_e = \kappa_0 T^{3/2} \;. \label{eq32}
\end{equation}
Here $\kappa_0 \sim 10^{-6} \; {\rm cgs}$ is Spitzer's thermal conductivity and the approximation ${\rm d} T/{\rm d} z \sim T/L$ is 
used for deriving equations~(\ref{eq30}) and (\ref{eq31}). These clearly indicate that the blackbody cooling becomes predominant in the preflare stage. 
Our model can be thus applicable to this stage. Note that, at the main burst stage, the blackbody cooling also mainly retains the thermal equilibrium of the system. 
\subsection{Microscopic Model of Baryonic Evaporation} 
We finally discuss the baryonic evaporation from the microscopic view point. In our model, the baryonic material is supposed to be heated 
by collisions between incident high-energy photons and the crustal matter. Considering that the ions are located in the center of 
Wigner-Seitz cell of the crust, the total internal energy per nuclei is naively given by the sum of Coulomb 
lattice energy $\varepsilon_{\rm lat} $ and thermal energy $\varepsilon_{\rm th}$ 
\begin{eqnarray}
 \varepsilon_{\rm tot}  \simeq && \varepsilon_{\rm lat} + \varepsilon_{\rm th} \nonumber \\ 
   = && 1.6\times 10^{-9} Z_{26}^{5/3} n_{{\rm cr},28}^{1/3} \nonumber \\
   && + 1.4\times 10^{-10} T_{{\rm cr},6} \ \ {\rm erg/nuclei} \;, \label{eq33}
\end{eqnarray}
(e.g., Shapiro \& Teukolsky 1983) where $Z_{26} = Z/26$ is the atomic number normalized by that of iron, 
$n_{{\rm cr},28} = n_{\rm cr}/10^{28}\ {\rm cm^{-3}}$ is the normalized crustal density, and $T_{{\rm cr} ,6}= T_{\rm cr}/10^6 $ K 
is the normalized crustal temperature. 

Since the lattice energy becomes predominant in the dense crustal surface, the total number of the evaporating nuclei $N$ should
satisfy the following equation; 
\begin{equation}
E_{\rm flare} \simeq N \varepsilon_{\rm lat} \simeq \int^{l}_{0}\ 4\pi R^2 n_{\rm cr}(z) \varepsilon_{\rm lat} \ \ dz \;, \label{eq34} 
\end{equation} 
where $l$ is the traveling depth of the incident photon through the magnetar crust, and $z$ is the depth from the magnetar surface. 

The hydrostatic balance between the surface gravity $g_{14} = g/10^{14}\ {\rm cm\ s^{-2}} $ and the pressure gradient force of 
a degenerate relativistic Fermi gas in the strong magnetic field $P_e \simeq \pi^2\hbar^2c^2 n_{\rm crust}^2/(3 eB )$ 
(TD95, Appendix A) gives
\begin{equation}
\frac{{\rm d}P_e}{{\rm d}z}  = \rho(z) g \simeq  m_p n_{\rm crust} (z) g \;,  \label{eq35}
\end{equation}
and leads to the density distribution of the crust, using the approximation ${\rm d}P_e/{\rm d}z \sim P_e/z$,
\begin{equation}
n_{\rm crust} (z)  \simeq  2.4\times 10^{31} B_{15} g_{14} \ z_3  \;, \label{eq36}
\end{equation}
where $z_3$ is the depth from the magnetar surface normalized by $10^3$ cm. 

Substituting the equation~(\ref{eq35}) into (\ref{eq34}), we can obtain the traveling depth of the photon 
through the crust;
\begin{equation}
l = 4.6\times 10^3 E_{{\rm flare}, 41}^{3/7} R_{6}^{-6/7} B_{15}^{-4/7} g_{14}^{-4/7}  \ {\rm cm} \;. \label{eq37} 
\end{equation}
The amount of the surface material heated by the collision between the high-energy photon and the crustal matter is, therefore, 
\begin{eqnarray}
M_{\rm ev} & = & 4\pi R^2 m_B \int^{l}_{0} n_{\rm crust}(z)\ dz \\ \nonumber 
                & = & 5.4\times 10^{24} R_{6}^{2/7} B_{15}^{-1/7}g_{14}^{-1/7} E_{{\rm flare},41}^{6/7} \ {\rm g} \;. \label{eq38}
\end{eqnarray} 
This is consistent with the value of equation~(\ref{eq16}) which is derived from the macroscopic view point in the framework of the 
solar-type magnetic reconnection model. Microscopic evaporation model would also support our magnetar flare model constructed 
in \S~2 and 3.  

Here we assume naively that all the released radiative energy is spent only for exciting ions, not for increasing their 
thermal and potential energies. Actually, it is not the easy task to clearly specify how the released radiative energy is distributed to 
each energy components during the complicated flaring activity. If we follow the energy equipartition law, a fraction of the 
released energy can be spent for exciting ions at least, and is enough for evaporating the baryonic matter with the mass 
$O(10^{24})$ g. In order to draw a physical picture for the energy distribution process more precisely, we need further study on the nature 
of strongly magnetized crystal (Harding \& Lai 2006). 

We would like to stress that the magnetic field stronger than $10^{14}$ G can modify structure of crystal of ions 
(e.g. Harding \& Lai 2006; Hansel et al. 2007). The binding energy of crystal in strong magnetic fields becomes different from 
that in weak fields. However, the structure of crystal in strong fields has not yet been studied well (see however, Usov et al. 1980). 
The main purpose of this paper is to construct a theoretical model for the magnetar's flaring activity from the macroscopic view 
point on the basis of the solar flare/CME model, not to establish a microscopic basis for our model. Thus, 
in this study, we give rough order-estimation of binding energy of crystal using a simple "Wigner-Seits" model in equation~(33).
\section{Summary}
According to the magnetic reconnection model which can correctly capture the solar flare/coronal mass ejection event, 
we propose a theoretical model for magnetar giant flares. It is highly suggestive that our model is accountable for the flaring 
activity associated with the giant flare from SGR 1806-20 consistently. Our main findings and characteristics of our model 
are summarized as follows: 

1. The temperature of the magnetar flare is essentially determined by the radiative cooling which compensates for the reconnection 
heating. The cooling effect due to the pair production partially contributes the thermal balance of the system, but is not significant. 
Since the blackbody cooling retains the thermal balance of the system, the flare temperature can be given by a simple scaling relation 
$T \propto R^{-1/4}\Delta t^{-1/4}B^{1/2} L^{1/2}$. 

2. During the flaring activity, the photon flux transports the released energy and heats the magnetar crust. Then the pressure of the 
crustal matter increases drastically and drives the evaporation flow. The incident radiative heat flux balances with the outgoing 
enthalpy flux of the evaporation flow. Neglecting the enthalpy contribution from the electron positron pair and the radiation field, 
the mass of the evaporated baryon can be represented by a scaling relation $M_{\rm ev} \propto R^2 T^3 \Delta t$. 

3. In our model, the preflare activity plays a role in supplying the baryonic matter into the magnetar's magnetosphere. 
The "prominence" which contains preflare-supplied baryonic matters is gradually uplifted via the energy injection from 
the magnetar's interior with maintaining a quasi-force-free equilibrium of the magnetosphere. Finally the prominence is 
erupted by the magnetic pressure force due to the loss of the equilibrium triggered by the magnetic reconnection at the 
main burst stage. The giant flare should be induced as the final outcome by the prominence eruption accompanied by large-scale field reconfigurations.

4. Our model predicts that the preflare activity produces a baryon-rich prominence. Then the erupted prominence is the origin of the observed 
radio-emitting ejecta associated with the giant flare from SGR 1806-20. In contrast, the post-flare loop formed in the main burst stage is polluted by 
dense baryonic matters and be trapped to the magnetar surface. This should be the origin of trapped fireball which causes 
the ultra-luminas $\gamma$-ray spike and the pulsating X-ray tail in the extraordinary flare observed from SGR 1806-20. 

\bigskip
We thank an anonymous referee whose helpful suggestions have helped us improve the paper. We also thank J. Matsumoto, H. Takahashi, S. Eguchi. and 
S. Mineshige for useful discussions. S.N is partially supported by Grants-in-Aid for Scientific Research of Japan through 19104006, 19740139 and 19047004. 
This work was supported by the Grant-in-Aid for the Global COE Program "The Next Generation of Physics, Spun from Universality and Emergence" from 
the Ministry of Education, Culture, Sports, Science and Technology (MEXT) of Japan.

%
%
%
%

\clearpage
\clearpage
\clearpage
\clearpage
\end{document}